\documentclass[twoside,
12pt,a4paper]{article}
\usepackage[marked,extnum,printedin]{myart2k}
\def\ifundefined#1{\expandafter\ifx\csname#1\endcsname\relax}
\providecommand{\comment}[1]{}

\usepackage{amsfonts}
\providecommand{\tthdump}[1]{#1}
\tthdump{}
\newcommand{\Cliff}[2][\comment]{{\ensuremath{%
\mathcal{C}\kern-0.18em\ell(#1,#2)}}}

\newcommand{\Space}[3][]{\ensuremath{\mathbb{#2}^{#3}_{#1}{}}}

\newcommand{\FSpace}[3][]{\ensuremath{ #2_{#3}^{#1}{}}}

\ifundefined{qed}
    \DeclareMathSymbol{\qed}{0}{AMSa}{"03}
\fi
\newcommand{\SL}[1][2]{\ensuremath{\FSpace{SL}{#1}(\Space{R}{})}}
\newcommand{\modulus}[2][]{\left| #2 \right|_{#1}{}}
\providecommand{\eqref}[1]{\textup{(\ref{#1})}}

\providecommand{\href}[2]{#2}
\iftth

\else

\fi

\ppnum{}{
\href{http://arXiv.org/abs/math-ph/0009013}%
{arXiv:math-ph/0009013}}%
{2000}

\titleshort{Nilpotent Groups and Clifford Analysis}
\authorshort{Vladimir V. Kisil}

\begin{document}
\title[Nilpotent Groups and Clifford Analysis]%
{Nilpotent Lie Groups \\
in Clifford Analysis and Mathematical Physics 
\\ \large
\emph{Five Directions for Research}}
\author[Vladimir V. Kisil]%
{\href{http://amsta.leeds.ac.uk/~kisilv/}{Vladimir V. Kisil}}

\address{%
School of Mathematics\\
University of Leeds\\
Leeds LS2\,9JT\\
UK}



\email{\href{mailto:kisilv@amsta.leeds.ac.uk}{kisilv@amsta.leeds.ac.uk}}

\urladdr{\href{http://amsta.leeds.ac.uk/~kisilv/}%
{http://amsta.leeds.ac.uk/\~{}kisilv/}}

\date{September 10, 2000}

\maketitle

\begin{abstract}
  The aim of the paper is to popularise nilpotent Lie groups (notably
  the Heisenberg group and alike) in the context of Clifford
  analysis and related models of mathematical physics. 
  It is argued that these groups are underinvestigated in
  comparison with other classical branches of analysis. We list
  five general directions which seem to be promising for further
  research. 
\end{abstract}  
\AMSMSC{30G35}{22E27, 43A85, 47A60, 47G, 81R30, 81S10} 


\section{Introduction}
\label{sec:introduction}
\epigraph{One simpleton can ask more
questions than a hundred of wise men find answers.}{}{}

The purpose of this short note is to advertise nilpotent Lie
groups among researchers in Clifford analysis. It is the author's
feeling that this interesting subject is a fertile area still waiting
for an appreciation. 

The r\^ole of symmetries in mathematics and physics is widely
acknowledged: the Erlangen program for geometries of F.~Klein and the
theory of relativity of A.~Einstein are probably the most famous
examples became common places already.
The field of Clifford analysis is not an exception in this sense:
the r\^ole of symmetries was appreciated from the very beginning, see
e.g. \cite{Cnops94a,Ryan95d} (to mention only very few papers). 
Moreover reach groups of symmetries are among corner stones which
distinguish analytic function theories from the rest of real
analysis~\cite{Kisil96c,Kisil97c,Kisil97a,Kisil01a}. This seems to be
widely accepted today: in the proceedings of the recent conference in
Ixtapa~\cite{Ixtapa99b} seven out of total seventeen contributions
explicitly investigate symmetries in Clifford analysis. But all of
these paper concerned with the semisimple group of \emph{M\"obius
(conformal) transformations} of Euclidean spaces. This group
generalises the $\SL$---the group of overwhelming importance in
mathematics in general~\cite{HoweTan92,Lang85} and complex analysis in
particular~\cite{Kisil97c,Kisil01a}.

On the other hand the M\"obius group is not the only group which may be
interesting in Clifford analysis. It
was argued~\cite{Howe80a,Howe80b} that the \emph{Heisenberg group} is
relevant in many diverse areas of mathematic and physics. The simplest
confirmations are that for example
\begin{itemize}
\item Differentiation $\frac{\partial}{\partial x}$ and
  multiplication by $x$---two basic operation of not only analysis
  but also of \emph{umbral calculus} in
  combinatorics~\cite{Cigler78,RomRota78}, for example---generate a
  representation of the Lie algebra of the Heisenberg group.
\item any quantum mechanical model gives a representation of the
  Heisenberg group.
\end{itemize}
Impressive continuation of the list can be found
in~\cite{Howe80a,Howe80b}.

Clifford analysis is not only a subfield of analysis but also has
reach and fertile connections with other branches: real harmonic
analysis~\cite{McIntosh95a}, several complex
variables~\cite{Gurlebeck-Malonek99,MiSha95}, operator
theory~\cite{JeffMcInt98a,Kisil95i}, quantum theory~\cite{Dirac67},
etc. Therefore it is naturally to ask about (cf.~\cite{Howe80a})
\begin{center}
  \emph{the r\^ole of the Heisenberg group in Clifford analysis}.
\end{center}

Unfortunately it was not written enough on the subject. The early
paper \cite{Kisil93c} just initiated the topic and the recent joint
paper \cite{CnopsKisil97a} only hinted about richness of a possible
theory. Thus the whole field seems to be unexplored till now. In order
to bring researchers' attention to the above question
we list in the next Section five (rather wide)
directions for future advances.


\section{Five Directions for Research}
\label{sec:open-problems}
In the joint paper~\cite{CnopsKisil97a} with Jan Cnops we constructed
two examples of spaces of monogenic functions generated by nilpotent
Lie groups. The first example is based on the Heisenberg group and
gives a monogenic space which is isometrically isomorphic to the
classic \emph{Segal-Bargmann space}
$\FSpace{F}{2}(\Space{C}{n},e^{-\modulus{z}^2}dz)$ of holomorphic
functions in $\Space{C}{n}$ square integrable with respect to the
Gaussian measure $e^{-\modulus{z}^2}dz$. The second example gives
monogenic space of Segal-Bargmann type generated by a Heisenberg-like
group with $n$ dimensional centre. The following propositions are
motivated by these examples. 
\begin{enumerate}
\item \textbf{Representation Theory}\\
  Representation theory of nilpotent Lie group by unitary operators in
  linear spaces over the field of complex numbers is completely
  described by the Kirillov theory of \emph{induced
  representations}~\cite{Kirillov62}. Particularly all irreducible
  unitary representations are induced by a character (one dimensional
  representation) of the centre of the group. Therefore the image of
  the centre is always one-dimensional. Representations in linear
  spaces with Clifford coefficients open a new possibility: unitary
  irreducible (in an appropriate sense) representations can have a
  multidimensional image of the group╢s
  centre~\cite{CnopsKisil97a}. Various aspects of such representations
  should be investigated.
\item \textbf{Harmonic Analysis}\\
  Clifford analysis technique is useful~\cite{McIntosh95a} for
  investigation of classic real harmonic analysis questions about
  \emph{singular integral operators}~\cite{Stein93}. Classical real
  analytical tools are closely related to the harmonic analysis on the
  Heisenberg and other nilpotent Lie
  groups~\cite{FollStein82,Howe80b,Stein93}. A combination of
  both---the Heisenberg group and Clifford algebras---could combine
  the power of two approaches in a single device.
\item \textbf{Operator Theory}\\
  The Segal-Bargmann space $\FSpace{F}{2}$ and associated
  orthogonal projection $P: \FSpace{F}{2} \rightarrow
  \FSpace{F}{2}$ produce an important class of \emph{Toeplitz
    operators} $T_a= P a(z,\bar{z}) I$~\cite{Coburn94a} which is a
  base for the \emph{Berezin
  quantisation}~\cite{Berezin88}. Connections between properties of an
  operator $P_a$ and its symbol $a(z,\bar{z})$ are the subject of
  important theory~\cite{CobXia94}. Moreover translations of Toeplitz
  operators to the Schr\"odinger representation gives interesting
  information on \emph{pseudodifferential operators} and their
  \emph{symbolic calculus}~\cite{Howe80b}. Monogenic space of the
  Segal-Bargmann type~\cite{CnopsKisil97a} could provide additional
  insights in these important relations.
\item \textbf{Functional Calculus and Spectrum}\\
  Functions of several operators can be defined by means of
  \emph{Weyl calculus}~\cite{Anderson69}, i.e. essentially using the
  Fourier transform and its connections with representations with
  the Heisenberg group~\cite{Kisil98a}. Another opportunity of a
  \emph{functional calculus} from Segal-Barg\-mann type spaces is also
  based on this group~\cite{Kisil98a}. On the other hand a
  functional calculus can be defined by the \emph{Cauchy formula}
  for monogenic functions~\cite{JeffMcInt98a,Kisil95i}. Simultaneous
  usage of monogenic functions and nilpotent groups could give a
  fuller picture for functional calculus of operators and associated
  notions of \emph{joint spectrums}.
\item \textbf{Quantum Mechanics}\\
  Observables of coordinates and momentums satisfy to the Heisenberg
  commutation relations $[p,q]=i\hbar I$, thus generated \emph{algebra
  of observables} representing the Heisenberg group. Even better: the
  representation theory of Heisenberg and other nilpotent Lie groups
  provides us with both non-relativistic \emph{classic and quantum}
  description of the world and a correspondence between
  them~\cite{Kisil96a,Kisil00a,Prezhdo-Kisil97}. On the other hand
  Clifford modules provide a natural description for \emph{spinor
  degrees of freedom} of particles or fields. Therefore a mixture of
  these two objects provides a natural model for quantum particles
  with spin. Yet such models and their advantages should be worked
  out.
\end{enumerate}
It should not be difficult to extend the list of open problems
(see the \hyperref[sec:introduction]{epigraph}).

\section{Acknowledgements}
\label{sec:acknowledgements}

This paper was prepared during author╢s research visit to University
of Aveiro, Portugal (August-September 2000) supported by the INTAS
grant 93--0322--Ext. I am grateful to the Prof.~Helmuth Malonek
for his hospitality and many useful discussions. 
\href{http://www.plmsc.psu.edu/~boris/}{Dr.~B.Veytsman} and
O.~Pilipenko gave me useful advises.

Any bibliography for a paper with a small size and a wide scope is
necessarily grossly incomplete. I appreciate an understanding of
readers who will not be disappointed if their relevant papers are not
listed among (almost random) references as well as an excuse for the
extensive self-citing.

{\small
\bibliographystyle{plain}
\bibliography{abbrevmr,akisil,analyse,aphysics,acombin}
}

\end{document}